\documentstyle[12pt]{article}

\input epsf.sty

\newcommand{\MS}{\overline{\rm MS}}
\newcommand{\g}{{\sl g}}

\begin{document}
\begin{titlepage}

\centerline{\large \bf Two-loop renormalization of Wilson loop
                       for Drell-Yan production.}

\vspace{10mm}

\centerline{\bf A.V. Belitsky\footnote{Alexander von
            Humboldt Fellow.}}

\vspace{10mm}

\centerline{\it Institut f\"ur Theoretische Physik, Universit\"at
                Regensburg}
\centerline{\it D-93040 Regensburg, Germany}
\centerline{\it Bogoliubov Laboratory of Theoretical Physics,
                Joint Institute for Nuclear Research}
\centerline{\it 141980, Dubna, Russia}

\vspace{20mm}

\centerline{\bf Abstract}

\hspace{0.8cm}

\noindent
We study the renormalization of the Wilson loop with a path
corresponding to the Drell-Yan lepton pair production in two-loop
approximation of perturbation theory. We establish the
renormalization group equation in next-to-leading order and
find a process specific anomalous dimension $\Gamma_{\rm DY}$
in the corresponding approximation.

\vspace{7cm}


\vspace{0.5cm}


\end{titlepage}


\noindent {\it 1. Introduction.} The increase of the experimental
accuracy necessitate the construction of the theory of power corrections
for hardronic reactions. For the time being there is a firm theoretical
ground only for the processes which admit the operator product expansion.
However, these are the processes without the latter which are of
the main interest. An important example of such a process is the
Drell-Yan (DY) lepton pair production with invariant mass $Q^2$. Recent
studies apparently reveal the correspondence between the soft-gluon
resummation which is necessary for the processes going near the
boundary of the phase space \cite{vanNeer84,Ste87,vanNeer89}
and the non-perturbative power corrections \cite{ConSte94,KorSte95}.
It was found that the ambiguity in the resummation of soft
gluons to the leading logarithmic accuracy manifest itself in the
power-like behaviour in the hard momentum scale leading to power
correction of the order $\Lambda_{1}/Q$ \cite{ConSte94}. It was argued
that this behaviour can be associated with particular matrix elements
of Wilson line operators \cite{KorSte95}. The nature of this
ambiguity was questioned in Ref.\ \cite{BenBra95} where the authors
attempted to identify the source of linear term in the perturbative
series for the Wilson loop corresponding to the DY production.
It was found that soft gluon resummation which allows to apply the
renormalization group methods and infrared renormalons are disconnected
since the former is related to the convergent analytical anomalous
dimensions while the latter enter as a boundary conditions to the
evolution equation. Moreover, it was established basing on the
resummation of the particular class of Feynman diagrams --- fermion
vacuum polarization bubbles --- that the leading ambiguity in the
perturbative series could be only $\Lambda_2/Q^2$. However,
the latter approach does not take into account the non-abelian nature
of QCD and, therefore, could miss some important features since it was
guessed that it is precisely the diagrams with three-gluon vertices
\cite{Kor96} which might be important in resolving these apparent
paradox. In the present paper we make a first step and evaluate the
Wilson loop in order ${\cal O}(\alpha_s^2)$. Later extending the result
to the case of nonzero gluon mass, $\lambda^2$, we will be able to
find the first non-analytical term in the expansion w.r.t. $\lambda$
which might be a signal for the first power correction \cite{Ben98}
up to limitations discussed below. However, the results given here are
of interest in their own right.

\begin{figure}[t]
\begin{center}
\vspace{8.5cm}
\hspace{-2.5cm}
\mbox{
\begin{picture}(0,220)(270,0)
\put(0,-30)                    {
\epsffile{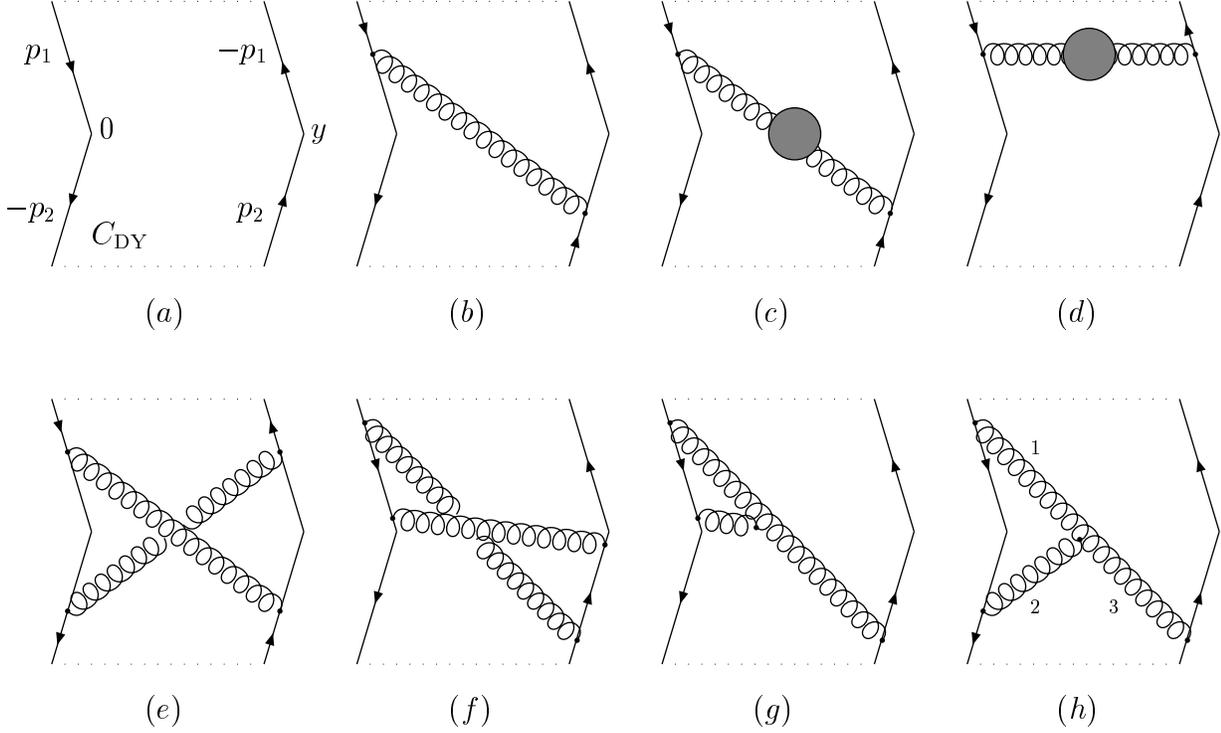}
                               }
\end{picture}
}
\end{center}
\vspace{-7.0cm}
\caption{\label{diagrams} Integration path in the Wilson loop in
Eq.\ (\protect\ref{WilsonLoop}) is shown in $(a)$. Non-zero one-loop
graph $(b)$. Different topologies of nonvanishing Feynman diagrams
contributing to the Wilson loop $W_{\rm DY}$ at order ${\cal O}
(\alpha_s^2)$ with maximal non-abelian factor in $(c)-(h)$. To
complete the set one has to add mirror symmetrical graphs which
can be taken into account by appropriate multiplicity factors,
$m_\sigma$, for graphs displayed here: $m_b = m_c = m_d = m_f = 2$,
$m_e = 1$, $m_g = m_h = 4$. Full blob stands for the sum of the
vacuum polarization bubbles due to fermions, gluons and ghosts.}
\end{figure}

\noindent {\it 2. Time-loop technique.} Following the approach of
Korchemsky and Marchesini we can express the soft part of the
factorized DY cross section \cite{KorMar93}
\begin{equation}
Q^2 \frac{d \sigma_{\rm DY} (z, Q^2)}{d Q^2}
= \sigma^{(0)}_{\rm DY} {\cal H}_{\rm DY} (Q^2) \widetilde W_{\rm DY} (z)
\end{equation}
via the Fourier transform, $\widetilde W_{\rm DY} (z)$, of the vacuum
average of the Wilson loop
\begin{equation}
\widetilde W_{\rm DY} (z) = \frac{Q}{2} \int_{- \infty}^{\infty}
\frac{d y_0}{2 \pi} e^{i y_0 \omega}
W_{\rm DY} (y), \qquad
W_{\rm DY} (y) = \frac{1}{N_c}
\langle 0|
{\rm Tr} {\cal T P}
\exp\left( i\g \oint_{C_{\rm DY}} dx_\mu B_\mu (x) \right)
|0 \rangle,
\label{WilsonLoop}
\end{equation}
with a path $C_{\rm DY}$ shown in Fig.\ \ref{diagrams}$(a)$. The
symbol ${\cal P}$ stands for path- while ${\cal T}$ for
(anti-)time-ordering to be explained below. Here $\omega \equiv
\frac{Q}{2} (1 - z)$ is a total soft gluon energy.

It is well known that a time-like cross section could not be
related to the imaginary part of any $T$-product of currents.
Contrary, it is given by the particular absorptive parts of the
Feynman diagrams. To be able to extract the imaginary part we
are interested in we can label in some way the field operators in
the amplitudes to the right and to the left of the cut. This can be
suitably done with the help of the Keldysh-like diagram technique
\cite{BalBra91,Bel97}. It allows to recast the program for the
calculation of the particular discontinuities of the Feynman diagrams
to the operator-like language. Consider, for instance, certain $S$-matrix
element which is given by the functional integral
${\cal M} = \int {\cal D} B \prod_j B_j
\exp\left( i \int dz {\cal L} \right)$.
The cross section of the process is
\begin{equation}
\label{GenerFunct}
\sigma =
{\cal M}^\dagger {\cal M} = \int \!{\cal D} B^{(-)} {\cal D} B^{(+)}
\prod_i B^{(-)}_i \prod_j B^{(+)}_j
\exp\left( i \int dz\ {{\cal L}^{(+)}}
- i \int dz\ {{\cal L}^{(-)}} \right).
\end{equation}
Here the "plus" and "minus" superscripts label the fields from the
direct and conjugated amplitudes, respectively. One can accept that
they are the components of a unique operator ${\cal B}_\mu =
\left( B^{(+)}_\mu, B^{(-)}_\mu \right)$ composed from the time-
and anti-time-ordered fields. Now the Green function for the
${\cal B}_\mu$ field is a $2\times 2$-matrix constructed from the
usual Feynman propagator, its conjugated analogue and its
discontinuity via the Cutkosky rules for the lines connecting the
direct and final amplitudes\footnote{By definition a $(-)$ field
stands to the left of $(+)$ ones.}. For the gluon propagator in the
Feynman gauge we have ($d = 4 - 2\epsilon$ is a space-time dimension)
\begin{equation}
\int d^d x e^{i qx}
\langle 0 |
{\cal T} \left\{ {\cal B}_\mu^a (x) \otimes {\cal B}_\nu^b (0)\right\}
| 0 \rangle
= \delta^{ab} g_{\mu\nu}
\left(
\begin{array}{cc}
- i D^{(++)} (q) & - i D^{(-+)} (q) \\
  i D^{(+-)} (q) &   i D^{(--)} (q)
\end{array}
\right),
\end{equation}
with the following form of the components\footnote{Here and below the
subscript ``$+$'' on the distribution, $\Delta = \delta,\theta$, means
the positivity of the energy flow through the cut: $\Delta_+ (q^2)
\equiv \Delta (q^2) \theta (q_0)$.} $D^{(++)} (q) = [D^{(--)} (q)]^\star
= (q^2 + i 0)^{-1}$, $D^{(-+)} (q) = - 2 \pi i \delta_+ (q^2)$.

Using these conventions we can easily write the Wilson loop
(\ref{WilsonLoop}) as follows
\begin{equation}
W_{\rm DY} (y) = \frac{1}{N_c}
\langle 0 |
{\rm Tr}
{\cal T}
\left\{
\Phi^{(-)}_{- p_1} [- \infty, y]
\Phi^{(-)}_{p_2} [y, - \infty]
\Phi^{(+)}_{- p_2} [- \infty, 0]
\Phi^{(+)}_{p_1} [0, - \infty]
\right\}
| 0 \rangle ,
\end{equation}
where ${\cal T}$ stands for the time-ordering for the $(+)$ and
anti-time one for $(-)$-fields and the Wilson lines are given by the
following expressions
\begin{eqnarray}
\Phi_{p_1} [y, - \infty]
&=& P \exp\left(
i \g \int_{- \infty}^{0}
d \sigma_1 p_{1 \mu} B_\mu (\sigma_1 p_1 + y)
\right), \nonumber\\
\Phi_{- p_2} [- \infty, y]
&=& P \exp\left(
- i \g \int_{0}^{- \infty}
d \sigma_2 p_{2 \mu} B_\mu (- \sigma_2 p_2 + y)
\right) .
\end{eqnarray}
Here the two quark momenta are defined in terms of the light-like
vectors, $p_1 = p_{1+} n^\ast$, $p_2 = p_{2-} n$, which fix
different tangents, i.e. $-$ and $+$ directions, on the light cone
and they are normalized according to $n^2 = n^{\ast 2} = 0$ and
$nn^\ast = 1$.

Let us note that the peculiarities of the renormalization of the Wilson
lines with a cusp \cite{Pol80,Ren80,KorRad87,And89} or/and with a path
lying on the light cone \cite{And89,Kor92} were studied before. It was
found that they require an additional renormalization and corresponding
anomalous dimensions play an important r\^ole in perturbative QCD. These
are the properties which we will extensively exploit presently.

Evaluating the one-loop graphs (Fig.\ \ref{diagrams}$(b)$ and mirror
conjugated) we can easily obtain unrenormalized expression
\begin{equation}
W^{(1)} = C_F \frac{\alpha_s}{\pi}
\frac{\Gamma (1 - \epsilon)}{\epsilon^2}
e^{-\epsilon \gamma_E} L^{\epsilon},
\end{equation}
where $L \equiv \left( - \frac{1}{4} y_0^2 \mu^2 e^{2\gamma_E}
+ i 0 \right)$ with $\mu^2$ being the $\MS$ scale parameter.
Subtracting the poles in $1/\epsilon$ we get well-known expression
\cite{KorMar93}
\begin{equation}
W_{\rm DY}^{(1)} (y) \equiv
{\cal R}_\epsilon W^{(1)} = C_F \frac{\alpha_s}{\pi}
\left\{ \frac{1}{2} \ln^2 L + \frac{\zeta (2)}{2} \right\}.
\end{equation}


\noindent {\it 3. Two-loop results.} Here we perform the evaluation
of the Wilson loop in the fourth order of perturbation theory. Expanding
the generating functional (\ref{GenerFunct}) in the perturbative
series we obtain a set of Feynman diagrams given in Fig.\ \ref{diagrams}.
Note that diagrams with virtual dressing of Wilson line vanish
identically in dimensional regularization since the integration over
loop momentum is given by a scaleless integral. Therefore, from several
number of cuts only those survive which are given by the product of
tree diagrams (up to exception of graph $(h)$). Moreover, on top of
this among the remaining graphs we will evaluate only those (they are
displayed in Fig.\ \ref{diagrams}) proportional to maximal non-abelian
and fermion components, i.e. $C_F C_A$ and $C_F T_F N_f$,
which is possible due to the non-abelian exponentiation theorem
\cite{Gat83}. According to it
\begin{equation}
W_{\rm DY} = 1 + \sum_{n = 1}^{\infty} W^{(n)}
= \exp\left( \sum_{n = 1}^{\infty} w^{(n)} \right),
\end{equation}
so that $W^{(2)} = \frac{1}{2} ( w^{(1)} )^2 + w^{(2)}$ and $W^{(1)}
= w^{(1)}$. In what follows we evaluate $w^{(2)}$.

\noindent {\it 3.1. Vacuum polarization diagrams.} The vacuum
polarization diagram in Fig.\ \ref{diagrams}$(c)$ reads
\begin{eqnarray}
w^{(2)}_{(c)} &=&
- C_F \g^2 {\cal S}_\epsilon \mu^{2\epsilon} p_{1+} p_{2-}
\int_{- \infty}^{0} d \sigma_2 \int_{- \infty}^{0} d \sigma_1 \nonumber\\
&\times& \int \frac{ d^d q}{(2 \pi)^d}
e^{- iq (\sigma_2 p_2 - \sigma_1 p_1 + y)}
D^{(--)} (q) D^{(++)} (q) 2 {\rm Im} \Pi_{+-} (q),
\end{eqnarray}
where ${\cal S}_\epsilon \equiv e^{\epsilon (\gamma_E - \ln 4 \pi )}$
and the imaginary part of the one-loop polarization operator is
\begin{equation}
{\rm Im} \Pi_{\mu \nu} (q)
= \alpha_s
\left[
\frac{C_A}{2} (5 - 3 \epsilon) - 2 T_F N_f (1 - \epsilon)
\right]
\frac{\Gamma (2 - \epsilon)}{\Gamma (4 - 2 \epsilon)}
\left( \frac{\mu^2 e^{\gamma_{E}}}{q^2} \right)^\epsilon
\theta_+ ( q^2 ) ( q^2 g_{\mu \nu} - q_\mu q_\nu ).
\end{equation}
After Fourier transformation with the help of formula ($x_\perp = 0$)
\begin{equation}
\label{formula1}
\int \frac{d^d q}{(2 \pi)^d} e^{- i qx }
\frac{\theta_+ (q^2)}{\left( q^2 \right)^n}
= \pi^{ - 1 - d/2 } 2^{ - 1 - 2 n }
\frac{\Gamma ( 1 - n ) \Gamma ( d/2 - n )}{
[ - 2 (x_- - i 0) (x_+ - i0) ]^{d/2 - n}}
\end{equation}
we get
\begin{equation}
w^{(2)}_{(c)}
= C_F \left( \frac{\alpha_s}{\pi} \right)^2
\left[
C_A (5 - 3 \epsilon) - 4 T_F N_f (1 - \epsilon)
\right]
\frac{\Gamma (1 - \epsilon) \Gamma (2 - \epsilon)
\Gamma (1 - 2 \epsilon) }{16 \epsilon^3 \Gamma (4 - 2 \epsilon)}
\left\{ 1 - \frac{\epsilon}{1 + \epsilon} \right\}
e^{- 2\epsilon\gamma_E} L^{2\epsilon}.
\label{VPc}
\end{equation}
The r\^ole of diagram $(d)$ is to cancel the second term in the curly
brackets in Eq.\ (\ref{VPc}) which comes from the ``gauge dependent''
$q_\mu q_\nu$-piece of the Landau gluon propagator. Multiplying these
contributions by appropriate multiplicity factors coming from adding
of analogous graphs with different attachments of gluon lines we obtain
\begin{equation}
w^{(2)}_\circ
= C_F \left( \frac{\alpha_s}{\pi} \right)^2
\left[
C_A (5 - 3 \epsilon) - 4 T_F N_f (1 - \epsilon)
\right]
\frac{\Gamma (1 - \epsilon) \Gamma (2 - \epsilon)
\Gamma (1 - 2 \epsilon) }{8 \epsilon^3 \Gamma (4 - 2 \epsilon)}
e^{- 2\epsilon\gamma_E} L^{2\epsilon}.
\end{equation}


\noindent {\it 3.2. Box-type diagrams.} Among a number of them most
vanish due to the light-like character of the paths and because
planar graphs are proportional to the fermion Casimir operator squared,
$C_F^2$, which is omitted due to non-abelian exponentiation theorem.
The evaluation of surviving graphs is the most straightforward in the
coordinate space where the propagators look like: $D^{(++)} (x) = -
\frac{i}{4} \pi^{- d/2} \Gamma (d/2 - 1)[- x^2 + i 0]^{1 - d/2}$,
and $D^{(-+)} (x) = - \frac{i}{4} \pi^{- d/2} \Gamma (d/2 - 1)[
- 2(x_- - i0)(x_+ - i0)]^{1 - d/2}$ (for $x_\perp = 0$). Fig.\
\ref{diagrams}$(e)$ gives\footnote{Note that the colour factor
of the box-type diagrams is $C_F \left( C_F - \frac{C_A}{2}\right)$
but we have kept only maximally non-abelian component.}
\begin{eqnarray}
w^{(2)}_{(e)}
&=& \frac{1}{2} C_F C_A \g^4
{\cal S}_{2\epsilon} \mu^{4\epsilon}
(p_{1+}p_{2-})^2
\int_{0}^{- \infty} d\sigma_1
\int_{- \infty}^{0} d\sigma_2
\int_{0}^{- \infty} d\sigma^\prime_2
\int_{- \infty}^{0} d\sigma^\prime_1 \\
&\times& D^{(-+)} (\sigma^\prime_2 p_2 - \sigma_1 p_1 + y)
D^{(-+)} (\sigma_2 p_2 - \sigma^\prime_1 p_1 + y)
= - C_F C_A \left( \frac{\alpha_s}{\pi} \right)^2
\frac{\Gamma^2 (1 - \epsilon)}{8 \epsilon^4}
e^{-2\epsilon\gamma_E} L^{2\epsilon}. \nonumber
\end{eqnarray}
Contribution of Fig.\ \ref{diagrams}$(f)$ is $W^{(2)}_{(f)} =
\frac{1}{4} W^{(2)}_{(e)}$. Assembling everything together we have
\begin{equation}
w^{(2)}_\Box
= - C_F C_A \left( \frac{\alpha_s}{\pi} \right)^2
\frac{3}{16}
\frac{\Gamma^2 (1 - \epsilon)}{\epsilon^4 }
e^{- 2\epsilon\gamma_E} L^{2\epsilon}.
\end{equation}


\noindent {\it 3.3. Non-abelian diagrams.} Finally, let us consider
the non-abelian diagrams. The typical contribution, for instance for
Fig.\ \ref{diagrams}(g) where the only contribution survives when the
loop is cut, looks like
\begin{eqnarray}
w^{(2)}_{(g)} &=& \frac{1}{2} C_F C_A \g^4
{\cal S}_{2\epsilon} \mu^{4\epsilon}
p_{1+} p_{2-}
\int_{- \infty}^{0} d \sigma_2
\int_{- \infty}^{0} d \sigma_1
\int_{- \infty}^{\sigma_1} d \sigma^\prime_1
\left(
\frac{\partial}{\partial\sigma^\prime_1}
-
\frac{\partial}{\partial\sigma_1}
\right) \nonumber\\
&\times&\int \frac{ d^d k}{(2 \pi)^d}
e^{- ik (\sigma_2 p_2 - \sigma^{\prime}_1 p_1 + y)}
D^{(--)} (k)
\int \frac{ d^d q}{(2 \pi)^d}
e^{- i (\sigma^{\prime}_1 - \sigma_1) qp_1 }
D^{(-+)} (q) D^{(-+)} (k - q).
\end{eqnarray}
The internal integral can be evaluated according to formula\footnote{Let
us add an interesting side-remark concerning the evaluation of this
integral. Actually, there is no need to perform its explicit calculation
with exponential weight. The dependence on $x$ can be fixed using the
properties of traceless tensors alone. Expanding the factor
$e^{-i q x}$ in Taylor series and using the fact that the tensor
$q_{\mu_1} q_{\mu_2} \dots q_{\mu_n}$ is traceless due to
presence of $\delta (q^2)$ in the integrand we can parametrize the
integral in terms of the only remaining vector $k$ according to
$\langle\langle q_{\mu_1} q_{\mu_2} \dots q_{\mu_n} \rangle\rangle =
\{ k_{\mu_1} k_{\mu_2} \dots k_{\mu_n} \} {\mit\Omega}$ (with
$\{ \dots \}$ standing for symmetrization and trace subtraction) and
perform tensor contraction with the help of the $d$-dimensional
generalization of Nachtmann's \cite{Nacht73} convolution
$\{ k_{\mu_1} k_{\mu_2} \dots k_{\mu_n} \} x_{\mu_1} x_{\mu_2}
\dots x_{\mu_n} = \frac{ \Gamma (n + 1) \Gamma (\lambda)}{\Gamma
(n + \lambda)} \left( \frac{k^2 x^2}{4} \right)^{n/2} C_n^{\lambda}
\left( \frac{i kx}{ \sqrt{- k^2 x^2}} \right)$ with $\lambda =
\frac{d}{2} - 1$ and $C_n^\lambda$ being the Gegenbauer polynomial
\cite{BatErd53V2}. Using third generating function for the latter
\cite{BatErd53V2} we can sum the series back. An $x$-free function
${\mit \Omega}$ can be easily evaluated then which results into
Eq.\ (\ref{formula2}).}
\begin{eqnarray}
\label{formula2}
\int \frac{d^d q}{(2 \pi)^d} e^{- i q x}
D^{(-+)} (q) D^{(-+)} (k - q)
&=& - \frac{\pi^\frac{d - 1}{2}}{\Gamma \left( \frac{d - 1}{2} \right)}
(4 \pi)^{2 - d} (k^2)^{\frac{d}{2} - 2} \theta_+ (k^2) \\
&\times&
e^{- \frac{i}{2} kx + \frac{i}{2} \sqrt{(kx)^2 - k^2 x^2}} \
{_1F_1}
\left(
\left.
{ \frac{d}{2} - 1
\atop d - 2 }
\right| - i \sqrt{(kx)^2 - k^2 x^2} \right), \nonumber
\end{eqnarray}
with $x^2 = 0$ and the second one with the help of Eq.\ (\ref{formula1})
making use of the integral representation of confluent hypergeometric
function ${_1F_1}$ \cite{BatErd53V1}.
Thus we get
\begin{equation}
\label{diagG}
w^{(2)}_{(g)}
= C_F C_A \left( \frac{\alpha_s}{\pi} \right)^2
\frac{\Gamma (1 - 2 \epsilon) \Gamma^2 (1 - \epsilon)
}{64 \epsilon^4 \Gamma (2 - 2 \epsilon)}
e^{- 2\epsilon\gamma_E} L^{2\epsilon}.
\end{equation}

Now let us observe a fact that the expressions for diagrams we have
considered so far with cut gluon Green functions coincide with
corresponding contributions evaluated with ordinary Feynman propagators.
This is not accidental but it is a mere consequence of the fact that their
collinear and cusp singularities --- the only ones we are interested in
--- appear when the gluons propagate along the light-cone directions or
they are all on short distances. In these cases cut and ordinary
propagators coincide.

There are two contributions coming from different cuts of diagram
$(h)$, namely, cut-I corresponding to the propagator combination
$D^{(++)}_1 D^{(++)}_2 D^{(-+)}_3$ and cut-II with
$D^{(-+)}_1 D^{(-+)}_2 D^{(--)}_3$. By the reason stated
above we have calculated instead the virtual graph which gives
\begin{equation}
\label{diagH}
w^{(2)}_{(h)}
= C_F C_A \left( \frac{\alpha_s}{\pi} \right)^2
\frac{\Gamma^2 (1 - \epsilon) }{64 \epsilon^4 (1 + \epsilon)}
\left\{
1 + \epsilon - 2 \epsilon
{F_2}
\left(
\left.
{ 1 , 1 + \epsilon , -2 \epsilon
\atop
2 + \epsilon , 1 - 2 \epsilon}
\right| 1,1 \right)
\right\}
e^{- 2\epsilon\gamma_E} L^{2\epsilon},
\end{equation}
with ${F_2}$ being Appel function \cite{BatErd53V1}.

Summing Eqs.\ (\ref{diagG},\ref{diagH}) multiplied by a factor of 4 we
obtain $w^{(2)}_\triangle$.


\noindent {\it 3.4. Counter-terms.} Before the subtraction of the
overall divergences we have to handle the sub-divergences. For this
reason we mention that although the diagrams with virtual subgraphs
(gluon vertex and propagator corrections) vanish for the case at
hand, corresponding counter-terms do not. Namely, combining the
contributions for uncut vacuum blobs and vertex functions results
into addendum
\begin{equation}
w^{(2)}_{\rm ct} = \frac{\alpha_s}{\pi} \frac{\beta_0}{4}
\frac{1}{\epsilon} W^{(1)},
\end{equation}
where $\beta_0 = \frac{4}{3} T_F N_f - \frac{11}{3} C_A$ is the first
expansion coefficients of the QCD $\beta$-function $\beta (\g)/ \g =
\frac{\alpha_s}{4 \pi} \beta_0$. This completes the set of
non-vanishing contributions we have to analyze.


\noindent {\it 4. Evolution equation in two-loop approximation.} Now
with the results derived in the previous sections we can find the
renormalized expression of the two-loop Wilson loop for the DY
production\footnote{Here we have used an expansion
${F_2} \left( \left. { 1 , 1 + \epsilon , -2 \epsilon
\atop 2 + \epsilon , 1 - 2 \epsilon} \right| 1,1 \right)
= - \frac{(1 + \epsilon)}{2 \epsilon}
\left( 1 - 2 \zeta (2) \epsilon^2 - 14 \zeta (3) \epsilon^3
+ {\cal O} (\epsilon^4) \right)$.}
($w^{(2)} = w^{(2)}_\circ + w^{(2)}_\Box + w^{(2)}_\triangle
+ w^{(2)}_{\rm ct}$)
\begin{equation}
\label{renormW_DY}
w^{(2)}_{\rm DY} (y) \equiv {\cal R}_\epsilon w^{(2)}
= \left( \frac{\alpha_s}{\pi} \right)^2
\left\{
w^{(2)}_3 \ln^3 L + w^{(2)}_2 \ln^2 L + w^{(2)}_1 \ln L
\right\},
\end{equation}
with
\begin{eqnarray}
&&w^{(2)}_3
= - \frac{1}{24} C_F \beta_0, \quad
w^{(2)}_2
= C_F C_A \left( \frac{67}{72} - \frac{\zeta (2)}{4} \right)
- \frac{5}{18} C_F T_F N_f, \nonumber\\
&&w^{(2)}_1
= C_F C_A \left( \frac{101}{54} - \frac{7}{4} \zeta (3) \right)
- \frac{14}{27} C_F T_F N_f.
\end{eqnarray}
Note the absence of $\ln^4 L$ terms \cite{vanNeer84,vanNeer89} due
to famous Lee-Nauenberg-Kinoshita theorem \cite{LNK}. Moreover, this
equation is in complete agreement with explicit two-loop calculation
of the ordinary Feynman graphs by van Neerven et al. \cite{vanNeer89}
(cf.\ Eq.\ (4.9) there adding charge renormalization counter-term)
provided we transform their result to the language of moments so that
we receive an addendum $\propto \zeta (2)$ which makes the results
coincide. The fact that the coefficient of the leading log, $w^{(2)}_3$,
is proportional to the $\beta$-function suggests that $W_{\rm DY}$
satisfies a renormalization group equation \cite{KorMar93,Kor92}. With
the expression (\ref{renormW_DY}) at hand it is easy to verify that
$W_{\rm DY} (y)$ indeed respects the following evolution equation
($\ln W_{\rm DY} (y) = w^{(1)}_{\rm DY} (y) + w^{(2)}_{\rm DY} (y)$)
\begin{equation}
\label{EvolEq}
\left(
\mu \frac{\partial}{\partial\mu}
+ \beta (\g) \frac{\partial}{\partial\g}
\right) \ln W_{\rm DY} (y)
= 2 \Gamma_{\rm cusp} (\g) \ln L + \Gamma_{\rm DY} (\g),
\end{equation}
with $\Gamma_{\rm cusp} = \left( \frac{\alpha_s}{\pi} \right)
C_F + \left( \frac{\alpha_s}{\pi} \right)^2 C_F \left( C_A \left(
\frac{67}{36} - \frac{\zeta (2)}{2} \right) - \frac{5}{9} T_F N_f
\right)$ being well-known universal cusp anomalous dimension
\cite{KorRad87} and a new process-dependent entry
\begin{equation}
\Gamma_{\rm DY} (\g) = \left( \frac{\alpha_s}{\pi} \right)^2
C_F \left(
C_A \left( \frac{101}{27} - \frac{7}{2} \zeta (3)
- \frac{11}{12} \zeta (2) \right)
+ \left( \frac{\zeta (2)}{3} - \frac{28}{27} \right) T_F N_f  \right) ,
\end{equation}
where $N_f$-dependent part is in agreement with calculation
of Ref.\ \cite{BenBra95}. An unusual feature of Eq.\ (\ref{EvolEq})
is that the ``anomalous dimension'' ${\bf\gamma} = 2 \Gamma_{\rm cusp}
\ln L + \Gamma_{\rm DY}$ depends explicitly on the renormalization
scale, $\mu$. This means the absence of the multiplicative
renormalizability of the light-like Wilson line.

Going to moments of the Fourier transformed Wilson loop,
$W_{\rm DY} \left( \frac{\mu N}{Q N_0} \right) = \int_{0}^{1}
dz z^{N - 1} \widetilde W_{\rm DY} (z)$, results to a mere
substitution of $y_0$ in the saddle point approximation by
$y_0 = - i 2 \frac{N}{Q}$. The solution of the evolution equation
(\ref{EvolEq}) is given by:
\begin{eqnarray}
W_{\rm DY} \left( \frac{\mu N}{Q N_0} , \alpha_s (\mu^2) \right)
&=& W_{\rm DY}
\left(1 , \alpha_s \left( Q^2 \frac{N_0^2}{N^2} \right) \right) \\
&\times&\exp
\int_{Q^2 \frac{N^2}{N_0^2} }^{\mu^2}
\frac{d \rho}{\rho}
\left(
\Gamma_{\rm cusp} \left( \alpha_s (\rho) \right)
\ln\left( \frac{\rho}{Q^2} \frac{N^2}{N_0^2} \right)
+
\frac{1}{2} \Gamma_{\rm DY} \left( \alpha_s (\rho) \right)
\right) , \nonumber
\end{eqnarray}
with $N_0 \equiv e^{- \gamma_E}$ and the upper limit $\mu$ which sets
the boundary for the maximal soft gluon energy.

The knowledge of $\Gamma_{\rm DY}$ at order ${\cal O} (\alpha_s^2)$
allows to predict DY inclusive production in next-to-next-to-leading
order approximation in large logarithms provided we have three-loop
expression for $\Gamma_{\rm cusp}$ which can be obtained by explicit
calculation of the Wilson line with a cusp or be extracted as a
coefficient in front of $1/[1-x]_+$-distribution of the three-loop
quark-quark splitting kernel once being evaluated in future.


\noindent {\it 5. Conclusion and outlook.} In present letter we
have calculated two-loop approximation of the Wilson loop for
the DY process. As we have mentioned in the introduction
it is straightforward to generalize present consideration by taking
into account the finite gluon mass propagator and look for
non-analyticity in $\lambda$ as a trace of power corrections.
This could be considered as an attempt to resolve the problem
with $1/Q$-power behaviour in DY reaction --- a question which has
attracted a lot of attention recently \cite{KorSte95,BenBra95,Ben98}.
But one has to be careful with this since even in the one-loop
calculations there is one-to-one correspondence between finite
gluon mass and renormalon based schemes \cite{Ben98} only for the
sufficiently inclusive quantities while they give different output
for observables with weighted final state, e.g. inclusive production
in $e^+e^-$-annihilation \cite{BenBraMag97}. More clear signal for
the leading power correction would be the first renormalon ambiguity
in the standard resummation of fermion bubble chains. But the
technique for handling the diagrams with non-abelian vertex still
has to be developed.

\vspace{0.5cm}

We deeply indebted to G.P. Korchemsky for numerous illuminating
discussions and to D.I. Kazakov for correspondence on a question
related to present study. The author was supported by the
Alexander von Humboldt Foundation and partially by Russian
Foundation for Fundamental Research, grant N 96-02-17631.

\end{document}